\documentclass[prd,nofootinbib,11pt]{revtex4}

\usepackage{hyperref}
\usepackage{graphicx}
\usepackage{amsmath,amssymb,amsfonts,amsthm,latexsym,stmaryrd}
\usepackage{marginnote}
\usepackage{color}
\usepackage{soul}
\usepackage{booktabs} 
\usepackage{siunitx} 
\newcommand{\beq}{\begin{equation}}
\newcommand{\eeq}{\end{equation}}

\newcommand{\U}{u}

\newcommand{\tb}{b}

\begin{document}

\title{Radial Oscillations of Neutron Stars with Vector-Induced Scalar Hair}
\author{Hamza Boumaza}
\affiliation{Laboratoire de Physique des Particules et Physique Statistique (LPPPS),\\
 Ecole Normale Supérieure-Kouba, B.P. 92, Vieux Kouba, 16050 Algiers, Algeria}
\date{\today}

\begin{abstract}
In this paper, we investigate the equilibrium configurations and radial perturbations of neutron stars within a subclass of gauge-invariant Scalar-Vector-Tensor (SVT) theories. By solving the generalized Tolman-Oppenheimer-Volkoff (TOV) equations for several values of the modified gravity parameter, we examine the impact of the vector-curvature coupling on the structure and properties of neutron stars. We then extend our analysis by deriving the quadratic action governing linear radial perturbations and computing both the normal modes associated with the matter sector and the scalar quasinormal modes arising from the additional propagating degree of freedom of the theory, which is able to propagate outside the neutron star. Our results show that the modified gravity parameter can significantly affect the mass-radius relation, the oscillation spectrum, and the stability properties of neutron stars, while preserving the coincidence between the onset of radial instability and the maximum-mass configuration, as in General Relativity.
\end{abstract}

\maketitle




\section{Introduction}
Black holes (BHs) and neutron stars (NSs) are among the most compact astrophysical objects  which provide strong-gravity environments where  possible deviations from general relativity could arise. These objects serve as  excellent astrophysical laboratories to  test gravity in strong-field regime and to probe the core of neutron stars \cite{capozziello2010beyond}. The detection data of gravitational waves sourced by the emergence of binary systems of black holes  and neutron stars \cite{abbott2016gw150914,LIGOScientific:2017vwq,LIGOScientific:2020zkf} open new windows for testing alternative theories of gravity. With the up coming high-precision observational gravitational waves will allow us to constrain, discriminate or exclude different subclasses of  alternative theories of gravity  as well as of matter models in the form of equation of state. 

The simplest extension of the Einstein’s theory of gravity is scalar-tensor theories in which a scalar field $\varphi$ in Einstein-Hilbert action, that couples nonminimally to the metric sector, is included. The most general scalar-tensor theory with equations of second order derivatives is called Horndenski's theory \cite{horndeski1974second}, and it has been extend to degenerate-higher-order-scalar-tensor (DHOST) theories with the aim of avoiding Ostrogradsky's instability \cite{Langlois:2015cwa}. An other alternative to modify gravity is by considering a vector field $A_\mu$ coupled to gravity, where the most general vector-tensor theories with second-order equations of motion are called generalized Proca theories \cite{Heisenberg:2014rta}. Recently, for the same aims a covariant scalar-vector-tensor theory of gravity was proposeded \cite{Heisenberg:2018acv}, by introducing coupling terms between vector field $A_\mu$ and the scalar field $\varphi$, where  Horndeski's theory and generalized Proca theories were unified in single action.   Due to the interactions between scalar and vector degrees of freedom, SVT theories and closely related vector-tensor and scalar-tensor extensions have important implications for cosmology and astrophysics, including black holes \cite{Heisenberg:2018mgr,Heisenberg:2017hwb}, neutron stars \cite{Lasky:2009sw,Kase:2020yhw}, and late-time cosmology \cite{Heisenberg:2018wye,Heisenberg:2018mxx,Kase:2018nwt,DeFelice:2016yws,BeltranJimenez:2013btb}. In particular, several results in vector-tensor theories, generalized Proca models, TeVeS-type constructions, and Horndeski-vector extensions can be viewed as limiting or related cases that share similar coupling structures and phenomenological features with SVT theories.

 Depending on whether the SVT theories are invariant with respect to the $U(1)$ gauge symmetry or not, it can be  classified into two cases. If $U(1)$ gauge symmetry is respected, the longitudinal component of a vector field vanishes, which left us with a scalar, two transverse vectors and two tensor polarizations). While by breaking of $U(1)$ gauge symmetry leads to an additional longitudinal scalar besides the five degree of freedoms. In the presence of a cubic-order scalar-vector interaction, new type of BHs solutions with scalar and vector hairs have been found for $U(1)$ gauge symmetric and shift symmetric (theory invariant under the transformation: $\varphi\rightarrow \varphi+const$)  theories,  where a scalar hair manifests  around the event horizon \cite{Heisenberg:2018vti}. These solutions show  stability against even-parity and odd-parity perturbations outside the event horizon under certain constraints \cite{Zhang:2024cbw, Heisenberg:2018mgr}.

Relativistic stars have been investigated within some specific subfamilies of generalized Vector-Tensor theories \cite{Sotani:2010re,Sotani:2010dr,Kase:2017egk,Chagoya:2017fyl} and of SVT theories \cite{Lasky:2008fs,LopezArmengol:2016irf,Sotani:2009nm}. For example, the authors of Ref.~\cite{Chagoya:2017fyl} considered a special case of vector Galileons with a one-parameter modification of the Einstein-Maxwell action, in order to study the influence of the vector field on the internal structure of relativistic stars. This model was later extended to more general cubic and quartic power-law derivative couplings in Ref.~\cite{Kase:2017egk}. In addition, relativistic charged stars in Einstein-Maxwell-scalar theories, which represent a particular subclass of SVT theories, were investigated in Ref.~\cite{Minamitsuji:2021vdb}, revealing the existence of new branches of solutions with nontrivial scalar configurations. In the present work, we investigate neutron stars within a subclass of $U(1)$ gauge-invariant and shift-symmetric Scalar-Vector-Tensor theories and compute several global properties that can be extracted from present and future observations, such as the mass, radius, and oscillation spectrum. In particular, we analyze the impact of the vector-curvature coupling on the equilibrium structure and radial stability of neutron stars, and compare our results with the corresponding predictions of General Relativity.

The study of radial perturbations provides an important tool to probe the internal structure and stability of compact stars, both in General Relativity \cite{unno1979nonradial,glass1983radial,Andersson:1995wu,Kokkotas:1986gd} and in modified gravity theories \cite{Kruger:2021yay,Mendes:2018qwo,Sotani:2004rq,Blazquez-Salcedo:2022klx}. In several modified gravity scenarios, additional gravitational degrees of freedom can propagate outside the star and generate characteristic quasinormal mode spectra. Such oscillation modes may carry important information about the nature of gravity in the strong-field regime and could potentially be constrained through gravitational-wave observations during the ringdown phase following compact-object mergers. In the framework considered in this work, the scalar sector gives rise to an additional propagating degree of freedom whose perturbations extend outside the neutron star and produce scalar quasinormal modes. Depending on the properties of the compact object and on the modified gravity coupling, these additional modes may induce observable deviations in the oscillation spectrum relative to the predictions of GR.

The organization of our paper is the following. In Section \ref{SEC1}, the general framework of SVT theories and specify the subclass under consideration are introduced. In Section \ref{SEC2}, we construct static and spherically symmetric neutron star backgrounds by solving the modified TOV equations. In Section \ref{SEC3}, the equation of motion for scalar and matter perturbations are derived from the expansion of the   second-order perturbation action.  Section \ref{SEC4} is devoted to the numerical computation of the fundamental  normal modes and of the impact the extra degree of freedom  on the characteristics of neutron stars in the proposed model. We conclude in Section \ref{conclusion} with a summary of our findings and discuss possible extensions of this work.
Additional details are provided in two appendices.

\section{Gravity models}\label{SEC1}

In this section, we briefly review the SVT gravity models  explored in this work.   The $U(1)$ gauge-invariant  SVT theories \cite{Heisenberg:2018acv} involve various  interactions between scalar field $\varphi$, vector field $A_\mu$  and the metric $g_{\mu\nu}$  through specific combinations. These combinations are principally the antisymmetric field strength tensor $F_{\mu\nu}$ and its dual $\tilde{F}_{\mu\nu}$
\begin{eqnarray}
    F_{\mu\nu}=\nabla_\mu A_\nu - \nabla_\nu A_\mu\quad\text{and}\quad \tilde{F}_{\mu\nu}=\frac{1}{2}\mathcal{E}^{\mu\nu\alpha\beta}F_{\alpha\beta},
\end{eqnarray}
where $\nabla_\mu$ is the covariant derivative and $\mathcal{E}^{\mu\nu\alpha\beta}$ is  anti-symmetric Levi-Civita tensor  satisfying the normalization $\mathcal{E}^{\mu\nu\alpha\beta}\mathcal{E}_{\mu\nu\alpha\beta}=-4!$. The  Lorentz-invariant combinations $F$, $\tilde{F}$ and $X$ are expressed in terms of $\nabla_\mu\varphi$, $F_{\mu\nu}$ and its dual $\tilde{F}_{\mu\nu}$ as 
\begin{eqnarray}
    F=-\frac{1}{4}F^{\mu\nu}F_{\mu\nu},\quad \tilde{F}=-\frac{1}{4}F^{\mu\nu}\tilde{F}_{\mu\nu},\quad Y=\nabla_\mu\varphi\nabla_\nu \varphi F^{\mu\alpha}F^{\nu}_{\;\:\alpha},  \quad\text{and}\quad\ X=-\frac{1}{2}\nabla_\mu\varphi \nabla^\mu\varphi.\nonumber\\
\end{eqnarray}
In order to keep the equation of motion up to second order, we need  to consider additional derivative interactions called   double dual Riemann tensor $L^{\mu\nu\alpha\beta}$ defined by
\begin{eqnarray}
   L^{\mu\nu\alpha\beta}=\frac{1}{4} \mathcal{E}^{\mu\nu\sigma_{1}\sigma_{2}}\mathcal{E}^{\alpha\beta\sigma_{3}\sigma_{4}}R_{\sigma_{1}\sigma_{2}\sigma_{3}\sigma_{4}},
\end{eqnarray}
 where $R_{\sigma_{1}\sigma_{2}\sigma_{3}\sigma_{4}}$ the Riemann tensor.

In the present work, we propose to investigate a sub-class of shift-symmetric and $U(1)$ gauge-invariant  SVT theories. Up to quadratic order,  the corresponding total action reads
\begin{equation}
S_{\rm grav}[g_{\mu\nu},\varphi]=\int d^4x\sqrt{-g}\left(K(X,F,\tilde{F},Y)+\sum_{i=3}^{i=4}(\mathcal{L}_{ST}^{(i)}+\mathcal{L}_{SVT}^{(i)})\right)\label{Stot},
\end{equation}
where $K$ is an arbitrary function of $X$, $F$, $\tilde{F}$ and $Y$.
The $\mathcal{L}_{SVT}^{(i)}$ and $\mathcal{L}_{ST}^{(i)}$, correspond to, respectively, SVT theories and Horndeski theories,  denote
the  elementary Lagrangians with second order derivatives of the scalar field, which are  explicitly   given by 
\begin{eqnarray*}
\mathcal{L}_{ST}^{(3)}&=&G_{3}(X)\nabla^{\mu}\nabla_{\mu}\varphi, \\
\mathcal{L}_{ST}^{(4)}&=&G_{4}(X)R+G_{4X}(X)\left((\nabla^{\mu}\nabla_{\mu}\varphi)^2-\nabla_{\mu}\nabla_{\nu}\varphi\nabla^{\mu}\nabla^{\nu}\varphi\right),\\
\mathcal{L}_{SVT}^{(3)}&=&\left(f_{3}(X)g_{\alpha\beta}+\tilde{f}_{3}(X)\nabla_{\alpha}\varphi\nabla_{\beta}\varphi\right)\tilde{F}^{\mu\alpha}\tilde{F}^{\nu\beta}\nabla_{\mu}\nabla_{\nu}\varphi, \\
\mathcal{L}_{SVT}^{(4)}&=&f_{4}(X)L^{\mu\nu\alpha\beta}F_{\mu\nu}F_{\alpha\beta}+\frac{1}{2}f_{4 X}(X)\tilde{F}^{\mu\nu}\tilde{F}^{\alpha\beta}\nabla_{\mu}\nabla_{\alpha}\varphi\nabla_{\nu}\nabla_{\beta}\varphi,
\end{eqnarray*}
where $R$ is the Ricci scalar associated with the metric $g_{\mu\nu}$ and the functions $f_i$ and  $G_i$  depend  on  kinetic term $X$. $f_{4 X}$ and  $G_{4 X}$ are the derivative of $f_{4 }$ and $G_{4 }$ with respect to $X$, respectively. In our paper, we will focus on the scalar-vector-tensor interactions  besides the Lagrangians Einstein-Hilbert  $R$ and the quintessence $X$, i.e. we choose $G_3$ and $G_4$ as follow
\begin{eqnarray}
    G_3=0,\quad G_4 =\frac{\kappa}{2}\quad \text{and}\quad K=X,
\end{eqnarray}
with $\kappa\equiv c^4/(8\pi G)$, being $G$ the Newton constant and $c$ the speed of light. For simplicity our study, we assume that   the functions $f_3$, $\tilde{f}_3$  and $f_4$ as constants
\begin{eqnarray}
 f_3=\beta_3,\quad\tilde{f}_3=0\quad\text{and}\quad f_4 =\beta_4.   
\end{eqnarray}
Therefore, the  action (\ref{Stot}) is reduced to
\begin{equation}
S_{\rm grav}=\int d^{4}x\, \sqrt{-g}\left(X+\frac{\kappa}{2}\,R+\beta_3 \tilde{F}^{\mu\alpha}\tilde{F}^{\nu}_{\;\:\alpha}\nabla_{\mu}\nabla_{\nu}\varphi+\beta_4L^{\mu\nu\alpha\beta}F_{\mu\nu}F_{\alpha\beta}\right)\,.
\label{S}
\end{equation}
 These choices  implies  that the propagation speed of  gravitational waves is strictly  equal to that of light \cite{Heisenberg:2018wye}. One can notice that the above model is described by two constants. The constant $\beta_3$ can be absorbed by the vector field as $A_\mu\rightarrow A_\mu/\sqrt{\beta_3}$ and the constant $\beta_4\rightarrow\beta_4\,\beta_3$, living us with a single parameter $\beta_4$. Varying the above action with respect to $\varphi$, yields
 \begin{eqnarray}
     \nabla_\mu\left[\nabla^\mu\varphi-\nabla_\nu\tilde{F}^{\mu\alpha}\tilde{F}^{\nu}_{\;\:\alpha}\right]=0.
 \end{eqnarray}
 By integrating  the action (\ref{S}) by part and using the above equation to eliminate $\nabla_\mu\nabla^\mu\varphi$, the new action is written as
\begin{equation}
S_{\rm grav}=\int d^{4}x\, \sqrt{-g}\left(\frac{\kappa}{2}\,R+\frac{1}{2} \varphi\nabla_{\mu}\nabla_{\nu}\left[\tilde{F}^{\mu\alpha}\tilde{F}^{\nu}_{\;\:\alpha}\right]+\beta_4L^{\mu\nu\alpha\beta}F_{\mu\nu}F_{\alpha\beta}\right)\,,
\label{Snew}
\end{equation}
where we observe that the scalar field play a role of Lagrange multiplier.
By integrating the action (\ref{S}) by parts and using the above equation to eliminate $\nabla_\mu\nabla^\mu\varphi$, the new action is written as
\begin{equation}
S_{\rm grav}=\int d^{4}x\, \sqrt{-g}\left(\frac{\kappa}{2}\,R+\frac{1}{2} \varphi\nabla_{\mu}\nabla_{\nu}\left[\tilde{F}^{\mu\alpha}\tilde{F}^{\nu}_{\;\:\alpha}\right]+\beta_4L^{\mu\nu\alpha\beta}F_{\mu\nu}F_{\alpha\beta}\right)\,,
\label{Snew}
\end{equation}
where we observe that the scalar field plays the role of a Lagrange multiplier. In this formulation, $\varphi$ does not carry independent dynamics, since it appears without a kinetic term, and therefore its variation imposes a constraint on the vector–curvature sector rather than generating an evolution equation. This type of structure is characteristic of auxiliary-field formulations in modified gravity, where scalar fields can enforce constraints instead of propagating additional degrees of freedom, as commonly encountered in $f(R)$-type theories and their scalar–tensor representations \cite{Sotiriou:2008rp}.  In this sense, the scalar field acts as a Lagrange multiplier selecting a restricted subset of admissible configurations in the vector–tensor sector, analogous to constrained constructions in generalized Proca and extended scalar–vector–tensor frameworks \cite{Heisenberg:2014rta,Heisenberg:2017hwb}. The action (\ref{Snew}) generates higher-order field equations, namely  a second-order  sixth-order partial differential equation for the tensor field (six degree of freedom), a second-order partial differential equation for the vector field (two degree of freedom), together with a sixth-order constraint equation obtained from the variation with respect to the scalar field. The system equations  have then seven degree of freedom.
\\
To study the gravitational action (\ref{S}) in the presence of a neutron stars, we must take into account an action representing a perfect fluid coupled minimally to the metric. Among the various  formulations of the matter actions proposed in the literature \cite{Taub:1954zz,schutz1970perfect,ray1972lagrangian,schutz1977variational,carter1989relativistic,brown1993action}, we consider the action proposed by Schutz ~\cite{schutz1970perfect}
\begin{eqnarray}
S_{\rm m} =\int d^4 x \sqrt{-g}\,P(\mu) , \label{action}
\end{eqnarray}
ignoring  the thermal effects of the matter. The pressure $P(\mu)$  is written   as a function of the chemical potential $\mu$,  which  is defined by the norm of the covector $\mu_\alpha$ (i.e. $\mu=\sqrt{-g^{\alpha\beta}\mu_\alpha\mu_\beta}$ with $u_\alpha$ is four-vector velocity), 
 \begin{equation}\label{decompositionofu}
     \mu_\alpha=\mu \, u_\alpha=\partial_\alpha L + A\partial_\alpha B\,,
 \end{equation}
 where $L$, $A$ and $B$ are three scalar fields. One can see that $\mu$ depends on the three scalar fields $L$, $A$ and $B$, trough the definition 
 \begin{eqnarray}
\mu= \sqrt{-g^{\alpha\beta}\left(\partial_\alpha L+ A\partial_\alpha B\right)\left(\partial_\beta L + A\partial_\beta B\right)}\,.\label{tmu22}
 \end{eqnarray}
In the special case of an   irrotational fluid, it  is sufficient to use a single scalar field by removing the scalar fields $A$ and $B$. By varying the action \eqref{action} with respect to these three scalar fields,  one  recovers the usual equations of motion for a perfect fluid. And the variation of the matter action with respect to $g_{\alpha\beta}$ must give the corresponding energy-momentum tensor,
 \begin{eqnarray}
 \label{T}
T_{\alpha\beta}=(P+\rho)u_\alpha u_\beta + P g_{\alpha\beta}\,,
\end{eqnarray}
 where $\rho$ is the  energy density.

\section{Background equations}\label{SEC2}
In order to describe the  background configuration of a non-rotating star, we consider a  static and  spherically symmetric metric given by 
\begin{equation}
ds^{2}=-f(r)dt^{2}+h(r)dr^2+r^2\left(d\theta^2+\sin^2\!\theta\,  d{\phi}^2\right)\,,
\label{ds}
\end{equation}
where $\{t,r,\theta,\phi\}$ are the time, radial, and angular coordinates, respectively, and the $f$ and $h$ are functions of  the radial coordinate. On this background, the scalar field $\varphi$ and the components of the vector field $A_\mu$ have the following ansatz
\beq
\label{phi}
\varphi = \varphi(r),\quad \text{and}\quad A_\mu = (A_0(r),A_1(r),0,0),
\eeq
obeying to the traceless condition,  as well as  avoiding the singularity of  the transverse component of vector field at $r=0$ \cite{DeFelice:2016cri}. Since the considering action (\ref{S}) is invariant under $U(1)$ gauge transformation, the radial component  $A_1(r)$ does not contribute to the vector-field dynamics. Therefore, we choose the gauge $A_1(r)=0$ in the following.\\ 
For the matter sector, the spatial components of the four-vector velocity of the fluid vanish, while the time component is derived from the normalisation condition $g_{\alpha\beta}u^\alpha u^\beta =-1$, yielding
\begin{eqnarray}
\label{u_bckgd}
\bar{u}_\alpha = \left\lbrace -\sqrt{f},0,0,0 \right\rbrace\,, 
\end{eqnarray}
showing an irrotational fluid and thus we can ignore the scalars $A$ and $B$. Assuming that the chemical potential $\mu$ is   a function of $r$, we integrate the component $t$ of Eq.(\ref{decompositionofu}) with respect to $t$. By doing so, we obtian
\begin{eqnarray}
\label{L_bckgd}
L(t,r)&=& -\sqrt{f}\mu t.
\end{eqnarray}
Substituting this result in the component $r$ of  Eq.(\ref{decompositionofu}), yeilds
\begin{eqnarray}\label{c}
\left(\mu\sqrt{f}\right)'=0.
\end{eqnarray}
 The matter conservation equation can be recovered by multiplying (\ref{c}) by $(dP/d{\mu})$ and using the formulas 
 \begin{eqnarray}\label{p+rho}
 P'=\mu'(dP/d{\mu})\quad \text{and}\quad \mu \,\frac{dP}{d\mu}=P+\rho.
 \end{eqnarray}
Alternatively, it can be  obtained from the $r$ component of the conservation of the energy-momentum tensor equation $\nabla_\mu T^\mu_\nu=0$, which gives
\begin{eqnarray}
P'=-\frac{f'(P+\rho)}{2f}\, ,
\label{e4}
\end{eqnarray}
where the pressure and the energy density are related trough the equation of state $P=P(\mu)$, or equivalently $P=P(\rho)$.

The substitution of the metric (\ref{ds}), scalar field  and vector field (\ref{phi}) into the total action $S=S_{\rm grav}+S_{\rm m}$ followed by an integrating by parts, yields
\begin{eqnarray}
S_{\rm bgd}&=&4\pi\int dr\,dt \left[\, r^2\sqrt{f h}\left(A_0'{}^2 \left(\frac{8 (h-1) \beta_4 + r\varphi '}{2f h^2 r^2}\right)+\frac{\kappa  \left(r h'+(h-1) h\right)}{h^2 r^2}-\frac{\varphi '^2}{2 h}+P\right)\nonumber\right.\\
&&\left.+ \lambda_0\, (\mu\sqrt{f})'\,\right]\,,
\label{S2}
\end{eqnarray}
where a prime denotes a derivative with respect to the radial coordinate $r$. The extra term  proportional to the Lagrange multiplier $\lambda_0$ is   added to  enforce  the constraint (\ref{c}). By varying the action (\ref{S2}) with respect to $f$ and $h$, we obtain the time and radial equations of motion which read, respectively,
\begin{eqnarray}
\frac{h'}{h}&=&\frac{1-h}{r}+\frac{r \varphi '^2}{2 \kappa }+\frac{A_0'^2 \left(8 (h-1) \beta_4 + r \varphi '\right)}{2 f h \kappa  r}+\frac{r  h \rho }{ \kappa }, \,\label{eh}\\
\frac{f'}{f}&=&\frac{h-1}{r}+\frac{r \varphi '^2}{2 \kappa }-\frac{A_0'^2 \left(8 (h-3) \beta_4 +3  r \varphi '\right)}{2 f h \kappa  r}+\frac{r  h P}{ \kappa },
\label{ef}
\end{eqnarray}
where, in the first equation, we have used 
\begin{eqnarray}\label{lamdaprime}
\lambda_0' =    r^2  \sqrt{h} \, \frac{dP}{d\mu}\,,
\end{eqnarray}
which follows from the variation of  \eqref{S2}  with respect to $\mu$ and the equation (\ref{p+rho}).

The scalar field and the vector field equations of motion are obtianed by varying  the action \eqref{S2} with respect to $\varphi$ and $A_0$, respectively. In the shift-symmetric and $U(1)$ gauge-invariant symmetry case, the conservation of  four-dimensional currents, $\nabla_\mu J^\mu_\varphi=0$ and $\nabla_\mu J^\mu_{A}=0$, are  reduced to 
\begin{eqnarray}\label{ephi}
\frac{d}{dr}\left(\sqrt{fh}\, r^2 J^r_\varphi\right)=0\quad\text{and}\quad \frac{d}{dr}\left(\sqrt{fh}\, r^2 J^r_A\right)=0,
\end{eqnarray}
with
\begin{eqnarray}
\label{Jr}
J^r_\varphi=\frac{\varphi '}{h}-\frac{ A_0'^2}{2 f h^2 r}\quad\text{and}\quad J_A^r=-\frac{A_0' \left(8 (h-1) \beta_4 + r \varphi '\right)}{f h^2 r^2}.
\end{eqnarray}
The singularity at the center is avoided, if we impose 
\beq\label{Jr0}
J^r_\varphi=0\quad\text{and}\quad J_A^r=0,
\eeq
which constitute a system of ordinary differential equations with a trivial solution given by $\varphi'=A_0'=0$.  In this limit, the equations of motion coincide with to those of general relativity. However, these equations also admit nontrivial solutions, given by
\begin{eqnarray}\label{Aphi}
    A_0'^2=16\beta_4 f (1-h) h \quad\text{and}\quad\varphi'=8\beta_4\frac{1-h }{ r},
\end{eqnarray}
which show that the constant  $\beta_4$ must be negative.  Replacing the above solutions in Eqs.(\ref{ef}-\ref{eh}), gives 
 \begin{eqnarray}
     \frac{f'}{f}&=&\frac{h-1}{r}+\frac{32 \left(1+2h-3 h^2\right) \beta_4 ^2}{\kappa r}+\frac{r h P}{\kappa },\label{df}\\
     \frac{h'}{h}&=& \frac{1-h}{r}+\frac{32 (1-h)^2 \beta_4 ^2}{\kappa r }+\frac{rh \rho  }{\kappa }.\label{dh}
 \end{eqnarray}
Here, the parameter $\beta_4$ quantifies the deviation from general relativity, with the GR equations are recovered in the limit $\beta_4\rightarrow 0$. 

In summary, $f$, $h$ and $P$ are the principal functions determined from the numerical integration of the radial differential equations system (\ref{e4}), (\ref{df}) and (\ref{dh})  for a chosen equation of state.

\section{Neutron stars profiles}
In this section, we model neutron stars as self-gravitating perfect fluids in thermal equilibrium, obeying a cold equation of state. We begin by examining the exterior region, then analyze the asymptotic behavior near the center of the star, and conclude with a full numerical integration across the radial coordinate.
\subsection{Expansions near the center of the star $ r = 0$ and spatial infinity $r\rightarrow\infty$}

Outside the star,  where $P = 0$ and $\rho=0$, equation \eqref{dh} can be integrated to yield
\begin{eqnarray}\label{exactsol}
 h=y^{-1}\left[\frac{r}{2M}\right],
\end{eqnarray}
where $M$ is the mass of the neutron star and $y^{-1}$ denotes the inverse of the function $y$, given by 
\begin{eqnarray}
y[h]=\frac{h}{h-1}\left(\frac{1}{h}+\frac{\beta_4^2}{\kappa}\frac{1-h}{h}\right)^{\frac{32\beta_4^2}{\kappa+32\beta_4^2}}.
\end{eqnarray}
To determine the event horizon,  we consider the limit $h\rightarrow \infty $, which leads to
\begin{eqnarray}
    r\rightarrow 2 M \,e^{\frac{32\beta_4^2}{\kappa+32\beta_4^2}\ln{\frac{\beta_4^2}{\kappa}}}.
\end{eqnarray}
The Schwarzschild radius, which equals  $2M$, is recovered in the limit $\beta_4\rightarrow 0$. This expression shows that, in our model, the event horizon is relatively shifted  to the Schwarzschild radius by a positive factor that depends on the coupling constant $\beta_4$. We note that in this solution, we recover also a similar singularity to Schwarzschild black hole at $r=0$ by taking the limit $r\rightarrow 0$. For large values of the radial coordinate $r$, the function $h$ behaves  as 
\begin{eqnarray}\label{hinf}
    h&\sim& 1+\frac{2 M}{r}+\frac{4 M^2 \left(\kappa -32 \beta _4^2\right)}{\kappa  r^2}.
\end{eqnarray}
Substituting this expansion into Eqs.(\ref{Aphi}) and (\ref{df}) , the
   second order expansions solutions of $f$, $A_0'$ and $\varphi'$ at spatial infinity are obtained as follows
\begin{eqnarray}
f&\sim& 1-\frac{2 M}{r}+ \frac{256 \beta _4^2 M^2 \left(\kappa -32 \beta _4^2\right)}{\kappa ^2 r^2},\\
A_0'^2&\sim& -\frac{32 \beta _4 M}{r}-\frac{64 \beta _4 M^2 \left(\kappa -32 \beta _4^2\right)}{\kappa  r^2},\\
\varphi'&\sim & -\frac{16 \beta _4 M}{r^2}.
\end{eqnarray}
Here, a nonvanishing scalar hair is induced by the coupling $\beta_4$, leading to modifications with respect to GR. If we integrate the expression of $A_0'$ with respect to  $r$, we find that the vector field exhibits an asymptotic behaviour proportional to $r^{1/2}$ at spatial infinity. However, the action and the equations of motion, in our model, are written in therms of the well behaved $A_0'$, since all physical quantities remain finite and well-defined.

Near the center of the star, we impose t regular boundary conditions $f'(0)=h'(0)=P'(0)=0$, which are  compatible with the following expansions 
\begin{eqnarray}
    f(r)=1+f_2 \,r^2,\quad h(r)=1+ h_2 \,r^2,\quad \rho(r)=\rho_c+ \rho_2 \,r^2\quad\text{and}\quad P(r)=P_c+P_2 \,r^2,
\end{eqnarray}
where $P_c$, $f_2$, $h_2$ and $P_2$ are constants.  One can determine  these constants by substituting the above  expressions into Eqs. (\ref{df}), (\ref{dh}) and (\ref{e4}), which enable us to find $f_2$, $h_2$ and $P_2$ in terms of $\rho_c$ and $P_c$. Near $r = 0$, we obtain
\begin{eqnarray}
 f(r)&=&1+\left(\frac{3 P_c+\rho _c}{6 \kappa }-\frac{64 \beta _4^2 \rho _c}{3 \kappa ^2}\right)\,r^2,\\
 h(r)&=&1+ \frac{\rho_c}{3\kappa} \,r^2\\
 P(r)&=&P_c-\left(\frac{\left(P_c+\rho _c\right) \left(3 P_c+\rho _c\right)}{12 \kappa }-\frac{32 \beta _4^2 \rho _c \left(P_c+\rho _c\right)}{3 \kappa ^2}\right) \,r^2.  \label{RC}
\end{eqnarray}
The function $h$ is not affected by the coupling constant $\beta_4$, but it modifies the profiles of $f$ and $P$. In the nonrelativistic regime $P_c\ll\rho_c$, the condition $P'(r)<0$ is satisfied only if the parameter $\beta_4$ lies within the range
\begin{eqnarray}
    \beta_4^2<\frac{3\kappa}{128}.
\end{eqnarray}
Finally, using Eq.(\ref{Aphi}), we find that, near the center, the scalar field and the vector field behave  as
\begin{eqnarray}
    \varphi'=-\frac{8 \beta _4 \rho _c}{3 \kappa }r,\quad\text{and}\quad A_0^2= -\frac{16 \beta _4  \rho _c}{3 \kappa }r^2,
\end{eqnarray}
which vanish for $\beta _4 =0$. It is worth noting that the above expansions around $r=0$ will serve as a boundary condition in the numerical analysis.
\subsection{Numerical integration of background equations}
In this subsection, we present  the numerical background solutions  of differential equations system governing the neutron star configurations, in order to show the physical characteristic of neutron star in all spacetime.

\begin{figure}[t]
\centering
\includegraphics[scale=0.8]{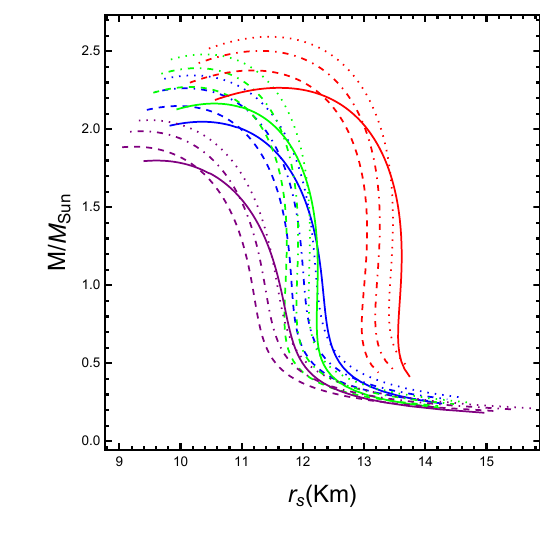} 
\includegraphics[scale=0.8]{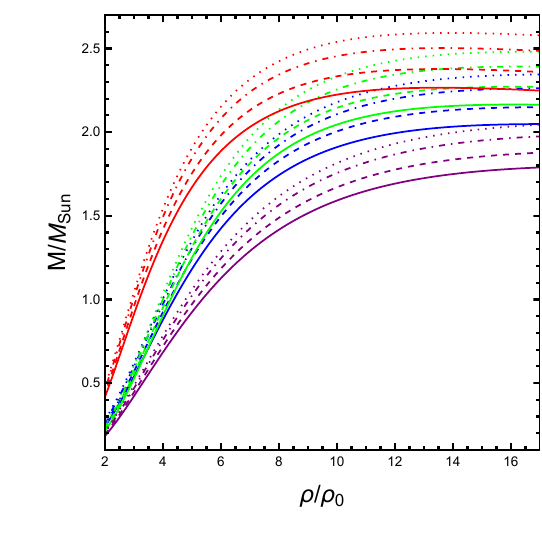} 
\caption{\small \small  
 Mass-radius ($M$-$R$, \textit{left panel}) and mass-central energy density ($M$-$\rho_c$, \textit{right panel}) relations for various values of the coupling parameter $\beta_4$. The curves correspond to $\beta_4 = 0$ (solid lines), $\beta_4 = -0.0033$ (dashed lines), $\beta_4 = -0.0046$ (dot-dashed lines), and $\beta_4 = -0.0053$ (dotted lines). Four different equations of state, distinguished by color, are considered: FPS (purple), SLy (blue), BSk20 (green), and BSk22 (red).}
\label{RM}
\end{figure}

In order to perform a numerical integration of  the (\ref{e4}), (\ref{df}) and (\ref{dh}),  we  consider four realistic equations of state, known as SLy, FPS, BSk20 and BSk22, discussed in  \cite{Haensel:2004nu,Potekhin:2013qqa,Pearson:2018tkr}, parametrised in the form
\begin{eqnarray}
\log_{10}\left(\frac{P}{{\rm g\;cm}^{-3}}\right) &=& \frac{\tb_1+\tb_2 \xi +\tb_3 \xi ^3}{1+\tb_4 \xi}\U\left[\tb_5 (\xi -\tb_6)\right]+(\tb_7+\tb_8 \xi)\U\left[\tb_9 (\tb_{10}-\xi )\right]+(\tb_{11}+\tb_{12} \xi)\U\left[\tb_{13} (\tb_{14}-\xi )\right]\nonumber\\
& & +(\tb_{15}+\tb_{16} \xi )\U\left[\tb_{17} (\tb_{18}-\xi )\right]
 +\frac{\tb_{19}}{\tb^2_{20} (\tb_{21}-\xi )^2+1}+\frac{\tb_{22}}{\tb^2_{23} (\tb_{24}-\xi )^2+1},\label{EOS}
\end{eqnarray} 
with 
\begin{eqnarray}
\U[x]\equiv\frac{1}{e^x+1}\,,\qquad  \xi=\log_{10}({\rho}/g\;{\rm cm}^{-3})\,.
\end{eqnarray}
Each equation of state is characterised by the values of the coefficients $\tb_i$. 
 For the  SLy and FPS equations of state, the coefficients are these coefficients are given by
 \beq
 \tb_i=a_i^{\rm HP} \quad {\rm for }\quad 1\leq i \leq 18\,,\qquad  \tb_j = 0\quad {\rm for} \quad 19\leq j\leq 24\,,
 \eeq
 where the  $a_i^{\rm HP}$ denote the coefficients $a_i$ of \cite{Haensel:2004nu}.
For the  BSk20 and BSk22 equations of state,  
\beq
 \tb_i=a_i^{\rm PFCPG} \quad {\rm for }\quad 1\leq i \leq 9\,,\qquad \tb_{10}= a_6^{\rm PFCPG}\,, \qquad   \tb_j =a_{j-1}^{\rm PFCPG} \quad {\rm for} \quad 11\leq j\leq 24\,,
 \eeq
 where the  $a_i^{\rm PFCPG}$ correspond to  the coefficients $a_i$ of 
\cite{Potekhin:2013qqa,Pearson:2018tkr}.

For different values of the parameter $\beta_4$ and of the central energy density $\rho_c$, the background equations are numerically integrated from the center to the surface of  the star.  At the center of the star, $r=0$, we impose the regularity  of the metric and energy density (see. Eq.\eqref{RC}). We then integrate from $r=0$ to the radius of the star $r=r_s$, which is determined by the condition $P(r_s)=0$ at the surface of the star. Imposing the continuity conditions, we perform a numerical integration from  $r=r_s$ to $r=\infty$, in practice $r=200 r_s$, allowing us to calculate the mass of the star, using equation \eqref{hinf}. Varying the central density from $2\rho_0$ to $18\rho_0$, we summarized our results  in Fig.\ref{RM}, for three values of the parameter $\beta_4$ and four different equations of state.

In Figure.\ref{RM}, the mass-radius relation for neutron stars, in our model, shows a deviation from general relativity. As the parameter $|\beta_4|$ increases, the maximum mass of the neutron stars increases with a slightly decreasing of the radius. This shows that our model allows for more massive and smaller radius neutron stars compared to GR. This behavior is reflected the enhanced effective pressure support within the star due to the scalar-vector coupling terms in the action \eqref{S}. This is an interesting property in the light of the recent astrophysical observation such as: the mass of the pulsar PSR J1614-2230 \cite{Demorest:2010bx} ($1.97 M_{Sun}$), or the mass of the compact object measured from the GW190814 event ($2.59 M_{Sun}$) \cite{LIGOScientific:2020zkf}. Furthermore, the results exhibit sensitivity to the equation of state, with BSk22 producing the highest masses and radii and FPS the lowest. These findings suggest that modified gravity models can reconcile soft nuclear equations of state with current astrophysical observations of high-mass neutron stars.

\section{Radial perturbations}\label{SEC3}
Now, we  turn our attention to the study of radial perturbations (see  Appendix~\ref{Appendix_axial} for more details). In order to investigate the radial spectrum predicted by our model, we  consider linear radial perturbations of the equilibrium stellar configurations described in previous section. The perturbed metric for a relativistic star, written in terms of the functions $\delta f(t,r)$, $\delta h(t,r)$ and $\delta w(t,r)$, reads  
\begin{eqnarray}
\label{perturbed_metric}
ds^2&=& -f (r) (1+\delta f(t,r))dt^2+h(r)(1+\delta h(t,r))dr^2+\sqrt{h(r)f(r)}\delta w (t,r)drdt\nonumber\\
&&+r^2\left(d\theta^2 + \sin^2 \theta d\phi^2\right),
\end{eqnarray}
while the perturbation of the scalar field and the vector field are given
by
\begin{eqnarray}
    \varphi=\varphi(r)+\delta\varphi(t,r)\quad\text{and}\quad A_\mu=(A_0+\delta A_0,\delta A_1,0,0).
\end{eqnarray}
 For the matter sector, the perturbed energy density, pressure, chemical potential, the scalar $L$ and  four-velocity  of the fluid are of the form 
\begin{eqnarray}
\label{4-velocity}
&\rho=\rho(r)+\delta\rho(t,r),\:\: P=P(r)+\delta P(t,r),\:\:\mu=\mu(r)+\delta\mu(t,r)+\delta^2\mu(t,r),\:\: \nonumber\\
& L=-\mu f(r)t+\delta L(t,r)\:\:\text{and}\:\:u^\alpha=\bar{u}^\alpha+\delta u^\alpha,
\end{eqnarray}
with
\begin{eqnarray}
\delta u^\mu = \left\lbrace \delta u^0(t,r),\delta u^1(t,r),0,0  \right\rbrace\,.
\end{eqnarray}
From the normalization condition $g_{\alpha\beta}u^\alpha u^\beta=-1$, the temporal component of the perturbed four-vector, up to second order, is calculated as 
\begin{eqnarray}\label{deltau0}
   \delta u^0(t,r)\sim \frac{\delta f(t,r)}{2 \sqrt{f(r)}}+\frac{1}{\sqrt{f(r)}}\left(\sqrt{h(r)}\delta u^1(t,r)  \delta w(t,r)+\frac{3}{8} \delta f(t,r)^2+\frac{1}{2}  h(r)\delta u^1(t,r)^2\right).
\end{eqnarray}
After substituting the perturbed metric \eqref{perturbed_metric}, the perturbed chemical potential and   the perturbed four-velocity \eqref{4-velocity} into the definition \eqref{decompositionofu}, we derive  the expression of $\delta ^2 \mu$ at second order in the perturbation 
\begin{eqnarray}
    \delta^2 \mu=-\frac{1}{2} h(r) \mu (r) \delta u^1(t,r)^2+\frac{1}{2} \delta f(t,r) \delta \mu (t,r)+\frac{1}{8} \mu (r) \delta f(t,r)^2.
\end{eqnarray}
In addition, the expressions for $\delta L'$ and $\delta \dot{L}$, where the dot denotes a derivative with respect to the temporal coordinate $t$, are derived at linear order in the perturbations, as 
\begin{eqnarray}
 \delta L'&=&\sqrt{h}\, \mu\, \delta w+h\, \mu\,\delta u^1,\\
 \delta \dot{L}&=& \frac{1}{2 }\sqrt{f} \,\mu\, \delta f(t,r)-\sqrt{f}\, \delta \mu ,
\end{eqnarray}
which implies the following equation
\begin{eqnarray}
   E_{\delta\mu}\equiv \mu\left(\delta \mu'- \frac{1}{2} \mu\, \delta \dot{f}+\frac{f'}{2 f}\delta \mu +\frac{\sqrt{h}\, \mu }{\sqrt{f}}\delta\dot{w}+\frac{h\, \mu  }{\sqrt{f}}\delta \dot{u}^1\right)=0,
\end{eqnarray}

If we multiply the above equation by $P_\mu$ and using the relations \eqref{p+rho}, we recover the radial component of matter conservation equation $\nabla_\alpha T^{\alpha\beta}$. Like in the background equation, this constraint is enforced by using  a Lagrange multiplier $\delta \lambda$, which added to the second-order perturbative expansion of the matter action. 
\begin{eqnarray}\label{Sm2}
   \delta^2 S_m &=&\int drdt\: \sqrt{fh}r^2\left(\frac{ (P+\rho )}{2 c_m^2 \mu ^2}\delta \mu ^2+\frac{\rho }{8}\delta f^2 -\frac{1}{2}  (\delta u^1)^2 h (P+\rho )+ \frac{ (P+\rho )}{2 \mu }\delta \mu \delta h-\frac{ P}{4}\delta f \delta h\right.\nonumber\\
   &&\left. -\frac{ P}{8}\delta h^2 +\frac{ P}{2}\delta w^2\right)+\delta\lambda \,E_{\delta\mu}\,,
\end{eqnarray}
where $c_m$ is the sound speed of the fluid. Varying this action with respect to $\delta \mu$ and  $\delta u^1$, gives

\begin{eqnarray}
    \delta \mu &=& -\frac{1}{2} c_m^2\, \mu  \left(\frac{2 }{\sqrt{f} \sqrt{h} r^2 (P+\rho )}\delta \lambda'+\delta h\right),\label{deltamu}\\
    \delta u^1 &=& \frac{1}{r^2 f \sqrt{h}}\frac{\delta \dot{\lambda}}{P+\rho },\label{deltau1}
\end{eqnarray}
respectively. In order to derive the dynamics of the scalar field and the Lagrangian multiplier associated with radial perturbations, we expand the total action up to second order in the perturbations.  After integrating by parts, we get
\begin{eqnarray}\label{sradial}
S_{\rm radial}&=&\int drdt \left[e_1\delta\dot{\varphi}^2 +e_2\delta\varphi'^2+q_1\delta\dot{\lambda}^2+q_2\delta\lambda'^2+q_0\delta\lambda^2+\delta\tilde{f}\left(a_0 V'+a_1 V+a_2\delta\lambda+a_3\delta\varphi'\right)\right.\nonumber\\
&&\left.+ \frac{2\sqrt{h}}{\sqrt{f}}\delta w \,\dot{V}+\delta F\,V'+a_4V^2+V(a_5\delta\lambda+a_6\delta\lambda'+a_7\delta\varphi')\right]\,,
\end{eqnarray}
with
\begin{eqnarray}
    V&=& \frac{\kappa  r\sqrt{f}  }{2 \sqrt{h}}\delta h+\frac{1}{2} \delta \lambda ,\\
    \delta F&=& \frac{1}{ \sqrt{h-1} \sqrt{-\beta_4 }} \left(\delta \dot{A}_0-\delta A_1'\right)-\frac{  \sqrt{h}}{\kappa  r\sqrt{f}}\delta \lambda,\\
    \delta\tilde{f}&=& \delta f+\frac{ \sqrt{h} }{\kappa  r \sqrt{f} }\delta \lambda-\frac{ 1}{2 \sqrt{-\beta_4}\sqrt{f} \sqrt{h}\sqrt{h-1} }\left(\delta \dot{A}_0-\delta A_1'\right),
\end{eqnarray}
where the coefficients $e_i$, $q_i$ and $a_i$ are background-dependant functions where their full expressions are provided in the appendix \ref{App1}. In the following, we choose   the gauge $\delta \tilde{f}=0$, which will allow us to decouple the matter sector from scalar field. The perturbations $\delta F$ and $\delta w$ appears as Lagrange multipliers where their equations of motion yield constraints. Varying the above action with respect to $ \delta w $ and $\delta F$, yields
\begin{eqnarray}
    \dot{V}=0\quad\text{and}\quad V'=0,
\end{eqnarray}
where theirs integration gives
\begin{eqnarray}
    V=c_0,
\end{eqnarray}
with $c_0 $ is a constant of integration.  This constant is omitted from the action (\ref{sradial}), since  it is  irrelevant to the dynamics of perturbations. Implementing these results in the  action (\ref{sradial}), it  is reduced to
 \begin{eqnarray}
     S_{\rm radial}&=&\int drdt \left[e_1\delta\dot{\varphi}^2 +e_2\delta\varphi'^2+q_1\delta\dot{\lambda}^2+q_2\delta\lambda'^2+q_0\delta\lambda^2\right]\,.\label{radial2}
 \end{eqnarray}
One  sees that the above action is separated into two fields, i.e. matter field and scalar field described by the functions $\delta\lambda$ and $\delta\varphi$, respectively. One also sees that the propagation speed  for the perturbed  scalar field is equal to that of light, while for the langrage multiplier is equal to the sound speed. Finally, it is  clear that there is no ghost or gradient instability in our model.

\section{Radial oscillations and numerical analysis}\label{SEC4}
In this section, we investigate the radial perturbations of neutron stars, focusing on the dynamics of both the scalar field and matter sector. In the previous section, we identify two decoupled propagating degrees of freedom, allowing us to analyze scalar field and matter perturbations independently.
\subsection{Matter perturbations}
\begin{figure}[t]
\centering
\includegraphics[width=14cm, height=8cm]{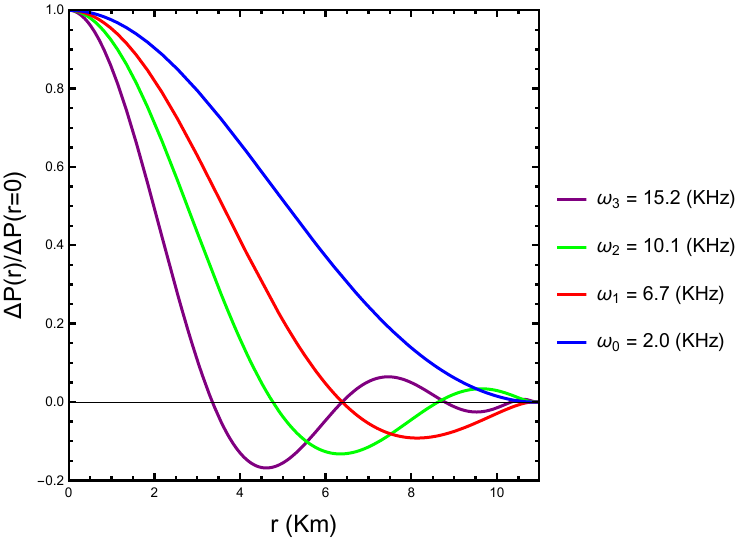}  
\caption{\small \small The first four  normalized Lagrangian perturbations of the pressure $\Delta P/P(0)$ as a function of the radial coordinate for $\beta_4=-0.00333$. The numerical solutions satisfy the boundary condition \eqref{con1}  and \eqref{con2}  in the case of a neutron star with SLy EoS and central density $\rho_c=10\rho_0$. 
}
\label{omega2}
\end{figure}

To study the radial oscillations of neutron stars within the framework of our gravitational model, it is necessary to compute the frequencies of their normal modes of oscillation. To this end, we introduce a small radial displacement from hydrostatic equilibrium at a given location $r$, denoted by  $\delta r (t,r)$. This radial displacement induces a corresponding Eulerian perturbation of the pressure, $\delta P (t,r)$, which is related to the Lagrangian perturbation $\Delta P$ through the relation
\begin{eqnarray}
 \Delta P= \delta P+   P'\;\delta r,
\end{eqnarray}
Furthermore, let us suppose that all perturbations have a harmonic time dependence of the form $\delta r(t,r) =\int e^{I\omega\, t}\delta\tilde{r}(\omega,r)\,d\omega$ and
$\Delta P(t,r) =\int e^{I\omega\, t}\Delta \tilde{P}(\omega,r)\,d\omega$, with $I$ is the imaginary number and $\omega$ is the characteristic frequency to be determined. For simplicity, we omit the tilde notation, but all perturbations quantities have to be understood as Fourier transforms.  Since $\delta u^1\equiv f^{-1}d\delta r/dt$, one can integrate equation \eqref{deltau1} with respect to the temporal coordinate $t$,  allowing us to express  $\delta r$ in terms of $\delta \lambda$. Then, implementing the obtained result in \eqref{deltamu}, and using the above relation, we obtain
\begin{eqnarray}\label{equdr}
    \frac{d}{dr}\left[\frac{\delta r}{r}\right]= \frac{1}{2} \frac{\delta r}{r}\left(\frac{f'}{f}-\frac{6 \kappa-64 \beta_4^2 (h^2-1) }{\kappa \,  r}\right)-\frac{\Delta P}{ \gamma \, r\,P },
\end{eqnarray}
where $\gamma = (1 + \rho/P) dP/d\rho$ is the adiabatic index at constant entropy. To insure that $(\frac{\delta r}{r})'$ is regular  at the center, we impose 
\begin{eqnarray}
   \Delta P= 3\gamma P \frac{\delta r}{r},\label{con1}
\end{eqnarray}
which corresponds to that we find in GR. Varying the action \eqref{radial2} with respect to $\delta \lambda$, gives
\begin{eqnarray}
\Delta P'&=&  \omega^2\frac{ \delta r}{r}\frac{r h(P+\rho )}{f} +\frac{\Delta P }{2 \gamma ^2 \kappa  P r}\left(32 \beta _4^2 (1-h) (3 h+1) \left(\frac{\partial\gamma}{\partial\rho}  (P+\rho )^2-\gamma ^2 P\right)\right.\nonumber\\
&&\left.+\frac{\partial\gamma}{\partial\rho}  (P+\rho )^2 \left((h-1) \kappa +h P r^2\right)-\gamma ^2 P \left(h \left(\kappa +r^2 (2 P+\rho )\right)-\kappa \right)+2 \gamma  \kappa  P r \gamma '\right) \nonumber\\
&&-\frac{ \delta r}{r}\frac{(P+\rho ) }{4 \kappa ^2 r}\left(7 \kappa ^2-h^2 \left(\kappa +P r^2\right)^2-2 h \kappa  \left(3 \kappa +P r^2\right)-64 \beta _4^2 \left(4 h^3 \rho  r^2\right.\right.\nonumber\\
&&\left.\left.+16 \beta _4^2 (h-1)^2 (h (9 h-2)+1)+(1-h) \left((h+1) (5 h+3) \kappa +(h-1) h P r^2\right)\right)\right).\label{eqdltap}
\end{eqnarray}
Similarly, we impose that $\Delta P/P$ is regular at the surface. Since, we have $P(r_s)=0$, the condition $\Delta P(r_s)/P(r_s)=0$ is ensured only if the coefficient proportional $\rho/P$ at $r=r_s$ vanish, which corresponds to  the quantity 
\begin{eqnarray}\label{con2}
   \Biggr[\frac{\Delta P}{P}-\frac{ \delta r}{r} \left(64 \beta _4^2 f (h-1) \left(16 \beta _4^2 (h-1) (h (9 h-2)+1)-(h+1) (5 h+3) \kappa \right)\right.\nonumber\\
  \left.+\kappa ^2 \left(f (h-1) (h+7)+4 h r^2 \omega ^2\right)\right)/\left(2 f (h-1) \kappa  \left(32 \beta _4^2 (3 h+1)-\kappa \right)\right)\Biggr]_{r=r_s}=0
\end{eqnarray}
In summary, the  linear system of equations \eqref{equdr} and \eqref{eqdltap} with the boundary conditions \eqref{con1} and \eqref{con2}  form a two points boundary value problems of the Sturm-Liouville type. After the static background equations are Solved, the numerical integration of these equations is carried out using the shooting method. We imposing  the conditions $\delta r = r$  and \eqref{con1} at the origin, and we perform a numerical integration to from the center to the surface. The constraint \eqref{con2} is  satisfied only for a discrete set of real values of $\omega$ corresponding to the normal modes.  In Fig.\ref{omega2}, we plot the perturbation $\Delta P$ for the first four eigenvalues $\omega_n$, where $ n = \{0, 1, 2,3,...\} $ represents the number of nodes inside the star, distinguishing with colors. In fact, the frequency corresponding to $n = 0$, called the fundamental mode, has no nodes between the center and the surface of the star and has the lowest frequency.  The first overtone ($n=1$) has a node, the second overtone ($n=2$) has two, and so on. 

\begin{figure}[t]
\centering
\includegraphics[width=14cm, height=10cm]{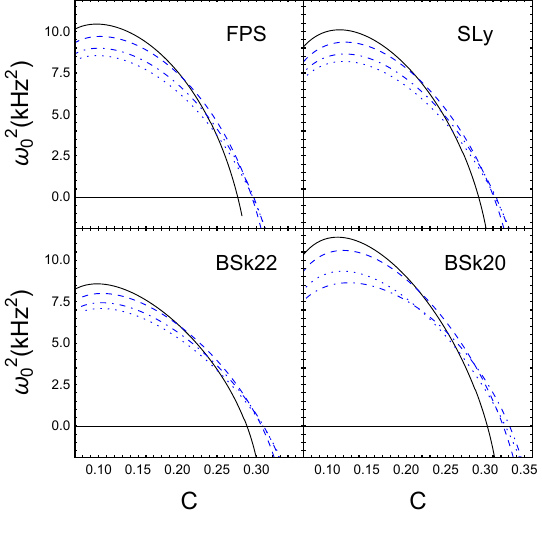}  
\caption{\small \small  
Squared frequency of the fundamental oscillation mode versus compactness, for the beyond-GR parameters $\beta_4=-0.00333$ (dashed lines),  $\beta_4=-0.00467$ (dotdashed lines) and $\beta_4=-0.00533$ (dotted lines),  together with  the GR ($\beta_4=0$) results (black lines). Four distinct equations of state are considered: FPS , SLy ,  BSk20  and BSk22. 
}
\label{omega3}
\end{figure}

We present in Fig.\ref{omega3} the squared frequency of the fundamental radial oscillation mode as a function of the stellar compactness
on the star compactness 
\begin{equation}
    C\equiv\frac{M}{r_s}\,,
\end{equation}
for four realistic equations of state: FPS, SLy, BSk20 and BSk22. Each panel shows the comparison between the predictions of General Relativity those of the scalar–vector–tensor theory for three  values of the coupling parameter $\beta_4$. As shown in the figure, the squared frequency  increases with compactness, reaches a maximum, and then decreases to zero at a critical energy density, where the neutron star becomes  unstable to radial perturbations. The parameter $\beta_4$  leads to a deviation from GR, which becomes increasingly significant in the high-compactness regime, modifying stability limit of neutron stars. These results confirm that $\beta_4$ acts as a stabilizing mechanism, as previously suggested in the context of   generalized Proca theories characterized by a $U(1)$-breaking vector field with derivative couplings \cite{Kase:2017egk}.  In other words, with the increase of $\beta_4$, the radial stability is indicated by a bigger central density value.

\subsection{Scalar field perturbations}
Now, we turn our attention to the scalar perturbation equation, obtained  by varying the action \eqref{sradial} with respect to $\delta\varphi$. Defining $\delta\varphi=\Phi/r$, the equation of scalar field reads
\begin{eqnarray}\label{pereq22}
  \frac{d^2 \Phi}{dr_*^2}+   \left(\omega ^2\; -\frac{ 1}{2 \; r}\frac{d}{dr}\left[\frac{f}{h}\right]\right)\Phi=0,
\end{eqnarray} 
where $\omega=\omega_R+i\,\omega_I$ is a complex quasinormal frequency where $\omega_R$ and $\omega_I$ are interpreted as the  oscillation frequency and the inverse of the damping time $\tau$, receptively. The variable $r_*$ is the tortoise coordinate defined as $r_*=\int dr\, \sqrt{h/f}$. Near the center of the star, the  general solution of the above equation behaves as\footnote{ The approximate solution \eqref{smalllambda} is obtained by  doing a Taylor expansion near the center of the star of the Eq.\eqref{pereq22}, which reads $$\Phi'' +\omega^2 \Phi=0\, .$$}
\begin{eqnarray}\label{smalllambda}
    \Phi\sim \Phi_-\, \cos(\omega\,r)+\Phi_+ \,\sin(\omega\,r)
    \qquad (r\to 0)
    \,,\end{eqnarray}
where $\Phi_+$ and $\Phi_-$ are constants of integration. 
Outside the star, the solution at spatial infinity is, like in GR, of the form
\begin{eqnarray}\label{solinf}
 \Phi\sim \Phi_{\rm in}e^{i \omega\, r_*}+\Phi_{\rm out}e^{-i \omega\, r_*}\qquad (r_*\to\infty)\,,
\end{eqnarray}
where $\Phi_{\rm in}$ and $\Phi_{\rm out}$ are also constants of integration. This solution is a linear combination of an outgoing and an ingoing wave. In order to have a purely outgoing wave and a regular perturbed scalar field solution at the center of the star, we have to set $\Phi_-=0$ and $\Phi_{\rm in}=0$. These boundary conditions are satisfied only for a discrete set of complex  values of $\omega$ corresponding to the  quasinormal modes. In practice, due to the exponential growth of the outgoing wave at spatial infinity, while the ingoing wave is small, we are not sure to have a purely outgoing wave function in the numerical treatment. To overcome this problem, we follow the method presented in Ref.~\cite{andersson1995new} 

\begin{figure}[t]
\centering
\includegraphics[width=14cm, height=10cm]{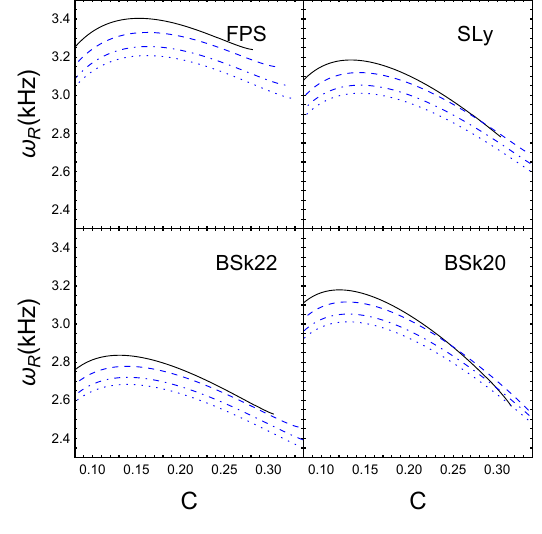}  
\caption{\small \small  
oscillation frequency as function of compactness, for the beyond-GR parameters $\beta_4=-0.00333$ (dashed lines),  $\beta_4=-0.00467$ (dotdashed lines) and $\beta_4=-0.00533$ (dotted lines),  together with  the GR ($\beta_4=0$) results (black lines). Four distinct equations of state are considered: FPS , SLy ,  BSk20  and BSk22. 
}
\label{wr}
\end{figure}

\begin{figure}[t]
\centering
\includegraphics[width=14cm, height=10cm]{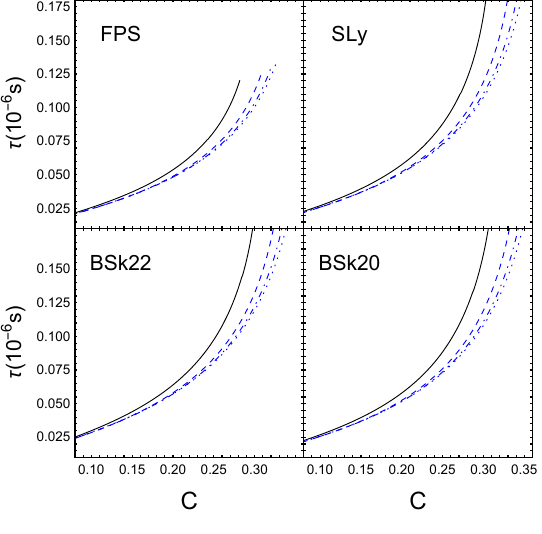}  
\caption{\small \small  
Imaginary part of $\omega$ ($\omega_I$) as function of compactness, for the beyond-GR parameters $\beta_4=-0.00333$ (dashed lines),  $\beta_4=-0.00467$ (dotdashed lines) and $\beta_4=-0.00533$ (dotted lines),  together with  the GR ($\beta_4=0$) results (black lines). Four distinct equations of state are considered: FPS , SLy ,  BSk20  and BSk22.}
\label{wI}
\end{figure}

In Fig.~\ref{wr} and Fig.~\ref{wI}, we have plotted the real part $\omega_R$  and the damping time $\tau=1/\omega_I$ of the fundamental  QNM, depending on the compactness of the neutron star.  Fig.~\ref{wr} reveals that the real part of  $\omega$ decreases  with compactness for all values if $\beta_4$. Meanwhile, we see in Fig.~\ref{wI} that, both in GR and in our model,  $\tau$ decreases as the neutron star mass increases.  In fact, as the neutron star becomes more compact, its oscillations not only slow down but also decay more rapidly. In addition, we  observe that the  real and imaginary parts is affected by the star's its equation of state. Among the four equations of state we have considered, the BSk22 EoS has the lowest oscillations and highest damping time.

\section{Conclusion}\label{conclusion}

In this paper, we have investigated possible deviations from General Relativity (GR) in the strong-field regime by studying the radial  perturbations of neutron stars within a specific subclass of SVT theories. As a starting point, we conducted both analytical and numerical analyses of static and spherically symmetric stellar configurations, solving the modified TOV equations for four realistic equations of state (EoS). By imposing vanishing scalar and vector currents, $J_\varphi=0$ and $J_A=0$, we derived an exact exterior solution, given by Eq.~\eqref{exactsol}, for the gravitational field outside the neutron star. To quantify deviations from GR, we constructed neutron star profiles across a range of central energy densities for various values of the SVT coupling parameter. Compared to their GR counterparts, neutron stars in this model can attain  higher masses and radii. 

We then investigated radial perturbations by expanding the total action to second order in the radial sector. By adopting an appropriate gauge, we identified two decoupled propagating degrees of freedom, corresponding to the scalar field and the matter sector, which allowed us to analyze their perturbations independently. Solving the resulting perturbation equations with suitable boundary conditions, we computed the normal modes for the matter sector and the quasinormal modes for the scalar sector. Our results show that the frequencies of both sectors, as well as the damping time of the fundamental scalar mode, exhibit deviations from GR predictions. These deviations depend on both the modified gravity parameter $\beta_4$ and the choice of the equation of state.

Finally, we note that our analysis has been limited to a simple one parameter a subfamily of VT theories. Future work should explore more general sectors within the SVT framework to determine whether other classes lead to distinct signatures. It would also be worthwhile to extend this study to include both axial and polar perturbations to assess the full phenomenological richness of SVT gravity in the context of neutron stars.

\appendix

\section{Coefficients in the gravitational equations of motion}\label{App1}
  The coefficients $a_i$, $e_i$ and $q_i$ in the action \eqref{sradial} are given by
  \begin{eqnarray}
    && q_0 =\left(c_m^2 (2 \kappa ^2 r^2 f''-f (h^2 r^2 (P+\rho ) (r^2 (P+2 \rho )-\kappa )-h \kappa  (4 \kappa -r^2 (P+\rho ))+4 \kappa ^2)\right.\nonumber\\
      &&\left. +32 \beta _4^2 f (-64 \beta _4^2 (h-1)^2 (h (3 h+8)+1)+(h-1) (h (3 h+5) P r^2\right.\nonumber\\
      &&\left.-  2 (h+1) (h+3) \kappa )+h (h (5 h+2)+1) \rho  r^2))+2 f h r^2 (P+\rho ) (h (64 \beta _4^2 (1-h)\right.\nonumber\\
      &&\left.+\kappa +P r^2)-\kappa  r (c_m^2)')\right)/\left(8 f^{3/2} \sqrt{h} \kappa ^2 r^4 (P+\rho )\right)\nonumber\\
      && q_1=-\frac{\sqrt{h}}{2r^2f^{\frac{3}{2}}(\rho+P)},\quad q_2=\frac{\sqrt{h}\,\,c_m^2}{2r^2\sqrt{f}(\rho+P)},\quad e_1=\frac{r^2\sqrt{h}}{2\sqrt{f}},\quad e_2=-\frac{r^2\sqrt{h}}{2\sqrt{f}},
      \nonumber\\
      && a_1 =\frac{128 \beta _4^2 (1-h) h}{\kappa  r},\quad a_2 =\frac{64 \beta _4^2 (h-1) h}{\kappa  r},\quad a_3=\frac{8 \beta _4 \sqrt{f} (1-h) r}{\sqrt{h}},\nonumber\\
      &&a_4=\frac{h^{3/2} \left(2 \left(192 \beta _4^2 (h-1)+\kappa +P r^2\right)-r^2 c_m^2 (P+\rho )\right)}{2 \sqrt{f} \kappa ^2 r^2},\nonumber\\
      && a_5=\frac{h^{3/2} \left(r^2 c_m^2 (P+\rho )-2 \left(64 \beta _4^2 (h-1)+\kappa +P r^2\right)\right)}{2 \sqrt{f} \kappa ^2 r^2},\nonumber\\
      && a_6= \frac{\sqrt{h}\,\, c_m^2}{\sqrt{f} \kappa  r},\quad a_7 = \frac{16 \beta _4 (h-1) }{\kappa }.
  \end{eqnarray}
 
\section{Even-Parity perturbations}
\label{Appendix_axial}
We consider linear perturbations of a static and spherically symmetric background spacetime, represented by  
\begin{equation}
    g_{\mu\nu} = g^{(0)}_{\mu\nu} + h_{\mu\nu},
\end{equation}
where \( g^{(0)}_{\mu\nu} \) denotes the background metric of the form \eqref{ds}, and \( h_{\mu\nu} \) is a small perturbation such that \( |h_{\mu\nu}| \ll |g^{(0)}_{\mu\nu}| \). Due to the spherical symmetry of the background, the perturbations can be classified into odd (axial) and even (polar) parity components, which decouple at the linear level.

We focus on the polar sector. The metric perturbations are expanded in terms of scalar spherical harmonics \( Y_{lm}(\theta,\phi) \), leading to the general decomposition:
\begin{align}
    &h_{tt} = \sum_{lm} \delta f^{lm}(r,t)\, Y_{lm}, \quad
    h_{tr} = \sum_{lm} \delta w^{lm}(r,t)\, Y_{lm}, \quad
    h_{rr} = \sum_{lm}  \delta h^{lm}(r,t)\, Y_{lm}, \quad
    h_{ta} = \sum_{lm} \beta^{lm}(r,t)\, \nabla_a Y_{lm},\nonumber \\
    &h_{ra}= \sum_{lm} \alpha^{lm}(r,t)\, \nabla_a Y_{lm}, \quad
    h_{ab} = \sum_{lm} K^{lm}(r,t)\, \gamma_{ab} Y_{lm} + G^{lm}(r,t)\, \nabla_a \nabla_b Y_{lm},
\end{align}
where \( \gamma_{ab} \) is the metric on the unit 2-sphere, and \( \{a,b\} = \{\theta,\phi\} \). The functions \( \delta f, \delta w, \delta h, \alpha, \beta, K, \) and \( G \) describe the radial and angular behavior of the perturbations.

We then consider an infinitesimal coordinate transformation \( x^\mu \rightarrow x^\mu + \xi^\mu \), with the gauge vector \( \xi^\mu \) restricted to even parity:
\begin{align}
    \xi^t = \sum_{lm} \xi_0^{lm}(r,t) Y_{lm},\quad \xi^r &= \sum_{lm} \xi_1^{lm}(r,t)  Y_{lm},\quad \xi^a = \sum_{lm} \xi^{lm}(r,t)\, \nabla^a Y_{lm}.
\end{align}
Under such a transformation, the metric perturbation functions transform accordingly, which allows one to eliminate gauge-dependent quantities. Here, we adopt the \textit{Regge–Wheeler gauge}, in which \( K=\beta^{lm} = G^{lm} = 0 \).  In this gauge, the polar part of the perturbed metric is described entirely by the functions \( \delta f, \delta w, \delta h\) and \( \alpha  \), which encode the physical content of the even-parity perturbations.

Finally, in the even-parity sector, the perturbation of the four-velocity \( \delta u^\mu \) can be decomposed in terms of the spherical harmonics as
\begin{eqnarray}
  \delta u^{a} =  \sum_{lm} \delta u^{a,lm}(r,t)\, \nabla^a Y_{lm},\quad
  \delta u^{t} =  \sum_{lm} \delta u^{0,lm}(r,t)\,  Y_{lm},\quad
  \delta u^{r} =  \sum_{lm} \delta u^{1,lm}(r,t)\,  Y_{lm}.
\end{eqnarray}
For the case $l=0$, $\delta u^{a}$ is vanished identically. Throughout this paper, the indices \( l \) and \( m \) have been omitted, since they do not affect  the results.

\bibliographystyle{JHEP}
\bibliography{DHOST_NS_biblio1}

\end{document}